\documentclass[aps,pra,superscriptaddress,showpacs,floatfix,amsmath,amssymb]{revtex4}
\usepackage{graphicx}
\begin{document}


\title{Multipartite Entanglement in a One-Dimensional Time Dependent Ising Model.  }

\author{Arul Lakshminarayan}
\email[]{arul@physics.iitm.ac.in}
\homepage[]{http://www.physics.iitm.ac.in/~arul}
\thanks{}
\affiliation{Department of Physics, Indian Institute of Technology Madras, 
Chennai, 600036, India.}
\author{V. Subrahmanyam}
\email[]{vmani@iitk.ac.in}
\thanks{}
\affiliation{Department of Physics, Indian Institute of Technology, 
Kanpur, 208016, India.}
\altaffiliation{}
\preprint{IITM/PH/TH/2004/7}

\date{\today}

\begin{abstract}
We study multipartite entanglement measures for a one-dimensional
Ising chain that is capable of showing both integrable and
nonintegrable behaviour.  This model includes the kicked transverse
Ising model, which we solve exactly using the Jordan-Wigner transform,
as well as nonintegrable and mixing regimes.  The cluster states arise
as a special case and we show that while one measure of entanglement
is large, another measure can be exponentially small, while
symmetrizing these states with respect to up and down spins, produces
those with large entanglement content uniformly. We also calculate
exactly some entanglement measures for the nontrivial but integrable
case of the kicked transverse Ising model. In the nonintegrable case
we begin on extensive numerical studies that shows that large
multipartite entanglement is accompanied by diminishing two-body
correlations, and that time averaged multipartite entanglement
measures can be enhanced in nonintegrable systems.

\end{abstract}


\maketitle
\newcommand{\newc}{\newcommand}
\newc{\beq}{\begin{equation}}
\newc{\eeq}{\end{equation}}
\newc{\kt}{\rangle}
\newc{\br}{\langle}
\newc{\beqa}{\begin{eqnarray}}
\newc{\eeqa}{\end{eqnarray}}
\newc{\pr}{\prime}
\newc{\longra}{\longrightarrow}
\newc{\ot}{\otimes}
\newc{\rarrow}{\rightarrow}
\newc{\h}{\hat}
\newc{\bom}{\boldmath}
\newc{\btd}{\bigtriangledown}
\newc{\al}{\alpha}
\newc{\be}{\beta}
\newc{\ld}{\lambda}
\newc{\sg}{\sigma}
\newc{\p}{\psi}
\newc{\eps}{\epsilon}
\newc{\om}{\omega}
\newc{\mb}{\mbox}
\newc{\tm}{\times}
\newc{\hu}{\hat{u}}
\newc{\hv}{\hat{v}}
\newc{\hk}{\hat{K}}
\newc{\ra}{\rightarrow}
\newc{\non}{\nonumber}
\newc{\ul}{\underline}
\newc{\hs}{\hspace}
\newc{\longla}{\longleftarrow}
\newc{\ts}{\textstyle}
\newc{\f}{\frac}
\newc{\df}{\dfrac}
\newc{\ovl}{\overline}
\newc{\bc}{\begin{center}}
\newc{\ec}{\end{center}}
\newc{\dg}{\dagger}

\section{Introduction}

The strictly quantum mechanical property of entanglement has attracted much
attention recently, mainly due to its role in quantum protocols such as 
teleportation, dense coding and other processes that involve transfer of quantum 
information. Entanglement has thus been thought of as a resource for quantum 
information processing, and perhaps quantum computing. While there is an 
understanding of what entanglement is, measures of the same are not so obvious, or 
well established.
 
Entanglement as quantum correlation has also been recently studied with the help 
of a slew of well-known models from condensed matter physics, such as the Ising 
and the Heisenberg models \cite{ConWoot,DenWoot,Sougato,Wang,Nielsen,Osterloh,Indrani}.
 Mainly, two-body correlations characterized by the concurrence \cite{Woot} have been studied
  in these systems. Also these were concerned mostly with stationary state properties, especially
  ground states. The entanglement content of a spin-chain, consisting of many spins, could be 
potentially much more than those that are present in two-body 
correlations, and nonstationary states are of potential interest in small chains, such as
those that may be realized in ion trap experiments.
 The difficulty is in defining proper measures of global entanglement content in such chains.
  Also it is important to note that much of the work has centered around those 
models that are completely integrable, mostly solvable by the Bethe Ansatz or by 
the Jordan-Wigner transform \cite{JordanWigner}.
 
Two-body or bipartite entanglement in pure states and
  its relation to chaos has been investigated more thoroughly mainly due to
   the von Neumann entropy of the reduced density matrices
    being an unambiguous measure of entanglement. One of the first works
    to find that chaos leads to larger entanglement production in this case,
    linked the classical Lyapunov exponent with the rate of
     entropy production \cite{MillerSarkar}.
     In this case it has been generally found that chaos encourages entanglement 
     \cite{Sakagami,Tanaka,Furuya,Lak,JayLak1,JayLak2},
and that complete chaos leads to an universal distribution of the eigenvalues of the
 reduced density matrices giving rise to an universal entanglement that depends
  only on the Hilbert space dimensions \cite{Lak,JayLak1}.
The first study that addressed the role of nonintegrability in many-body entanglement used
 the Harper model \cite{LakSub}, while later works used the quantum baker map \cite{ScottCaves}, the Frenkel-Kontorova model \cite{Hu}
 and disordered spin chains \cite{Santos}. 

While the relation between entanglement and chaos or nonintegrability is subtle even
 in bipartite systems, it gets even more so in the case of many-body systems.
  It has been claimed that opposite effects have been observed in this case,
   namely a decrease of entanglement with
chaos \cite{Santos}. However in the case of one-particle states it has been observed that
 the average of all the two-body correlations present in the system does increase with chaos
 \cite{LakSub,Hu}, while near-neighbour correlations decreases with chaos,
  where the nearness of the neighbour depends on a kind of quantum correlation
   length \cite{LakSub}. Thus it would seem that chaos
in these cases can encourage distant entanglement, even of a two-body type. However most  
studies have addressed two-body entanglements, and not global or multipartite 
entanglements. The exceptions are recent works of Scott and Caves \cite{ScottCaves} that make use of  
a measure due to Meyer and Wallach \cite{MeyWall}, called here the $Q$ measure, and indeed show,
 using the examples of a quantum kicked rotor and the quantum baker map,
that an increase in chaos entails larger global entanglement.
 
 Admittedly, global measures 
of entanglement are only now beginning to be explored and it is likely that the various
proposed  measures quantify different aspects of entanglement in multipartite states,
aspects that need further elucidation. We now briefly recapitulate the definitions of
 the different entanglement measures used in this paper. We emphasize that we will 
 throughout this paper deal exclusively with {\em pure} states.
 
\begin{description} \item {\bf 1. Concurrence}: 
 The concurrence in two qubits $i$ and $j$ that
 are in the joint state $\rho_{ij}$ is given by the following procedure \cite{Woot}:
  calculate the eigenvalues of the matrix $\rho_{ij}\,\tilde{\rho_{ij}}$, where $\tilde{\rho_{ij}}
  =\, \sigma^y\otimes \sigma^y \rho_{ij}^*
 \sigma^y\otimes \sigma^y$, and the complex conjugation is done in the standard computational
 basis. The eigenvalues are positive and
  when arranged in decreasing order if they are $\{\ld_1,\ld_2,\ld_3,\ld_4\}$,
   the concurrence is $C_{i,j}=\mbox{max}\left (\sqrt{\ld_1}-
   \sqrt{\ld_2}-\sqrt{\ld_3}-\sqrt{\ld_4},\,0\right)$. This is such that $0\le C_{i,j}\le 1$,
   with the concurrence vanishing for unentangled states and reaches unity for maximally
    entangled ones. The entanglement of formation of the two qubits is known to be
    a monotonic function of $C_{i,j}$ and hence concurrence is itself a good measure
    of entanglement. In the case of many-qubit pure states, we study concurrence 
    between any two qubits by tracing out the others qubits, and studying the resultant 
    density matrix. Thus this is a ``two-body'' correlation. 
    It is known that in typical states multipartite entanglement is shared among many qubits rather than in a pairwise
    manner \cite{ScottCaves}. We use the following two measures to study multipartite entanglement.  
 \item {\bf 2. Residual tangle and the $N$-Tangle}: If two qubits $i$ and $j$ are in a pure state
  $|\psi\kt$, the
  concurrence reduces to $|\br \psi |\sigma^y \otimes \sigma^y |\psi^*\kt|$. The state $|\psi^*\kt$ is such
 that its components in the computational basis are the complex conjugates of those of 
 $|\psi\kt$.
  It was found that the square of the concurrence $\tau_{i,j}=C_{i,j}^2$ is a more natural
   measure and is now
   called the tangle \cite{CoffValWoot}.  We can also define the tangle between one spin
    (say the $k$-th)
   and the rest of the spins if the overall state is pure. This is because in this case,
    the Schmidt decomposition gives two unique eigendirections to the rest of the spins
     corresponding to those
  eigenvalues of the reduced density matrix that are nonzero. There will be utmost only two 
  such values as the nonzero eigenvalues of the two parts are identical. Thus the 
  rest of the qubits can also be effectively thought of as a two-state system. 
  The tangle between spin $k$ and the rest, the one-tangle, is $\tau_{k,(\mbox{rest})}= 4 \det(\rho_k)$, where $\rho_k$ is 
  the reduced density matrix of the $k$-th qubit (we will also call this simply $\tau_k$, 
  not to be confused with the $n$-tangle introduced below).
   This was used to define a purely three-way entanglement measure in a pure state of three qubits
  as 
  \beq
  \label{restang}
   \tau_{1,(23)}-\tau_{1,2}-\tau_{1,3}.
  \eeq This quantity, called the residual tangle, is independent
   of the focus qubit, in the above this being the qubit numbered 1 \cite{CoffValWoot}, and hence
   stakes its claim as a pure three-way entanglement measure. 
  The construction used here to define the residual tangle was generalized to $N$ qubits in
   Ref. \cite{WongChrist}, effectively defining a measure of multipartite entanglement,
    the $n$-tangle as $\tau_N= |\br \psi |\sigma^{y\, \otimes N} |\psi^*\kt|^2$. This is
     evidently the tangle for $N=2$, for $N=3$ this is the residual tangle, while for $N>3$ and 
    odd this vanishes. Thus this measure is used only for $N$ even, with the exception of
    $N=3$. It has been shown to be an entanglement monotone \cite{WongChrist} and hence is a candidate for
    measuring multipartite entanglement.
    It is maximal (unity) for GHZ type states, but can also be maximum for states such as the
     product state of two groups of four spins in the 4-GHZ states. It is of course zero for
      completely unentangled states. It must be noted that the direct generalization of the 
      residual tangle in Eq.~(\ref{restang}) is conjectured to be positive \cite{CoffValWoot},
       and is not the same as the $n$-tangle for $n>3$.

    \item{\bf 3. The Meyer and Wallach $Q$ measure:}
    The geometric multipartite entanglement measure $Q$ \cite{MeyWall}, has been shown to be simply related
     to one-qubit purities \cite{Brennen}, which makes their calculation and interpretation straightforward.
      It also seems to have the potential for being experimentally measurable. This is defined as:
  \beq
  Q(\psi)=2\left(1-\f{1}{L}\sum_{k=1}^{L}\mbox{Tr}(\rho_k^2)\right).
  \eeq
 From the unit trace of density matrices, it follows
   immediately that for qubits $1-\mbox{Tr}(\rho_k^2)=2 \det(\rho_k)$.    
 Thus we get that 
 \beq
 Q(\psi)=\f{1}{L}\sum_{k=1}^L \tau_k.
 \eeq 
 This measure is therefore simply the average of the tangle between a given qubit and 
 the rest, averaged over this ``focus'' qubit. The relationship between $Q$ and single
 spin reduced density matrix purities has lead to a generalization of this measure to higher
 dimensional systems and taking various other bipartite splits of the chain \cite{Scott}.
    In some ways, for many states, the $n$-tangle and the $Q$ measure seem to be measuring
quite ``orthogonal" aspects of entanglement, as we see below, eventhough we can and do construct
GHZ type of states that maximize both these measures.
\end{description} 

  It needs to be stressed that it is not true that
   nonintegrability in general produces more entanglement for arbitrary states. 
   There are very simple operators, trivially integrable ones,
    that can create maximally entangled states out of particular
   unentangled initial states. However, apart from being true only for particular
   initial states, the entanglement will oscillate in time and can be completely destroyed
    once again. An example is provided by the Hamiltonian $H= S^x_A \otimes S^x_B$ of two spin-half 
    particles. With $\hbar=1$ we get,
    \beq
    \exp(-i J S^x_A \otimes S^x_B t )|11 \kt = \cos(Jt/4)|11\kt -i \sin(Jt/4)|00\kt
    \eeq
    where the states are in the standard $S^z$ diagonal basis and $|1\kt$ is the state with
    eigenvalue $1/2$.
     The 2-tangle is simple to calculate and is  
     \beq
     \tau = |\br \psi |\sigma_A^y \otimes \sigma_B^y |\psi^*\kt|^2=\sin^2( J t/2),
     \eeq
     which follows on substituting the above state. The quantity $\tau$ is unity for maximally
     entangled states and is zero for unentangled states.     
    Thus after a time $t=\pi/J$, the spins will be maximally entangled,
    whereas after twice that time they would be totally unentangled once more.
    Generalizations of such Hamiltonians, and states, to larger spins and
    to larger number of qubits also yield similar results, and is elaborated upon later below.
     
    In this paper we study issues related to entanglement sharing in spin chains that 
    can range from the integrable to the nonintegrable, but which  nevertheless involve
     only nearest neighbour interactions and
    are translationally invariant. This is in contrast to models
    which have been studied so far, which are essentially single body dynamics, such as the 
    Harper or the quantum baker map, that have been mapped onto many qubit systems by 
    means of an isomorphism of the Hilbert space. This implies that the interactions
    need not be nearest neighbour and can in fact involve all to all interactions. 
    In particular the model we study is a kicked Ising model of which the 
    kicked transverse Ising model is a special case. The kicked transverse Ising model is 
    integrable and we solve it using the Jordan-Wigner transform, and thereby study the 
    entanglement generated by this evolution. The zero-field version of this is trivially
    solvable and a class of states that follow in this case have been previously studied as the
    ``cluster states'' \cite{Cluster}. We show that while the cluster states have large entanglement as measured
    by one entanglement measure, $Q$ measure, it has an 
    exponentially small (in number of spins) $n-$tangle.
    We also show how symmetrizing the states produces those that have large entanglement 
    according to both these measures. We also emphasize that the kicking is unlikely to be a
    crucial aspect for the issues discussed here, and on the contrary is more suitable for 
    implementations in say ion-trap experiments.
     
    We numerically study the nonintegrable case and compare it with the integrable one.
    We find that while nonintegrability does discourage two-body entanglement, multipartite entanglement
    is increased on the average. In fact this entanglement comes at the cost of two-body
    correlations. Again, nonintegrability is not required to produce maximally entangled
    states, but produces states that retain large entanglement without disentangling. Thus
    we study the time averaged entanglement measures starting with the vacuum (all spin down)
    state and study it as a function of the strength and tilt of the external field. It is
    seen that the parameter space corresponding to nonintegrable chains is capable of 
    having substantial entanglement.

    \section{The Kicked Ising Model}
    
    The model with which we principally study these issues in this paper is a variant 
    \cite{Prosen}
    of the transverse Ising model, a variant that is at once both dynamically interesting 
    and easier to implement with present day quantum technologies. The usual transverse 
    Ising model has been studied in the context
    of both entanglement and state transport. It is an intriguing model that is integrable
    due to a mapping via the Jordan-Wigner transformation, from interacting spins to a
    collection of noninteracting spinless fermions. The relevance of this model
     to many physical systems has long been appreciated and it is a well studied model,
     with a quantum phase transition separating ferromagnetic and paramagnetic phases at zero 
     temperature as a parameter is varied. The Hamiltonian for $L$ spin $1/2$ particles is 
     \beq
     H_I = J \sum_{n=1}^{L}S^x_n S^x_{n+1} + B \sum_{n=1}^{L}S^z_n,
     \eeq
     where $J$ is the local exchange coupling strength and $B$ an external transversal field.
     For $J>2B$, the system is in a ferromagnetic phase with nonzero expectation values of the
     $S^x$ component of the spin, while for $J<2B$ the system is paramagnetic with vanishing
     $S^x$ spin expectation value, the point $J=2B$ being a quantum critical point. 
     
     The variant mentioned above involves applying a {\em tilted}
      external field
      impulsively at regular intervals of time \cite{Prosen}. The operator that evolves states from one
      application of the field to the next is the quantum map or propagator whose
      spectral properties determine the time evolution. The Hamiltonian is
      \beq
      H=J \sum_{n=1}^{L}S^x_n S_{n+1}^x +
       B \sum_{k=-\infty}^{\infty}\delta\left(k-\f{t}{T}\right) \,  \sum_{n=1}^{L}\left(
        \sin(\theta)S_{n}^z + \cos(\theta) S_n^x \right),
       \eeq
     while the unitary quantum map is (the time $T$ between the kicks sets the time scale and
       is set to unity):
      \beq
      U=\exp \left(-i  J \sum_{n=1}^{L}S_{n}^x S_{n+1}^x \right)
       \,\exp\left(-i B \sum_{n=1}^{L}\left(\cos(\theta) S^x_n + \sin(\theta) S^z_{n}\right) 
       \right).
      \eeq
      When the field is  transverse  ($\theta=\pi/2$), due to the noncommutativity
       of the components of the spin operator, the above is not
      equal to $\exp(-iH_I)$, and gives rise to genuinely different dynamics.
       However it has been shown that this ``kicked'' transverse Ising model
        is integrable \cite{Prosen} and there are suggestions to show that it also undergoes
	 a quantum phase transition and belongs to the same
       universality class as the usual transverse Ising model \cite{Milburn}.
        In this integrable model too the key is the Jordan-Wigner transformation, and 
	we solve the problem exactly as opposed to the assumption of the thermodynamic
	limit in Ref. \cite{Milburn}. For 
	$0 \,< \theta\, <\pi/2$ it appears that the model is {\em nonintegrable} and
	capable of showing mixing behaviour in the thermodynamic limit \cite{Prosen}.

       Define the following unitary operators: 
       \beqa
        U_{aa}(J_a)= \prod_{n=1}^{L_0}\exp \left(-i J_a S^a_{n} S^a_{n+1}\right),\,\\
	U_{x,z}(B,\theta)=\prod_{n=1}^{L}\exp\left(-i B \left(\cos(\theta) S^x_n + \sin(\theta) S^z_{n}\right)
	\right).
       \eeqa
     Here $L_0=L$ for periodic boundary conditions and is $L-1$ for open chains, and $\theta$
     is an angle of tilt of the magnetic field in the $x-z$ plane. The letter $a$ can be $x$
      or $z$.
    For the most part we will consider the operator
     \beq
     U =  U_{xx}(J_x)\, U_{x,z}(B,\theta). 
     \eeq
This series of unitaries are quantum gates on nearest neighbour pairs of qubits
 and on individual qubits. Ion-trap quantum computing provides one way of implementing the 
 above. The two-qubit operator $U_{xx}$ maybe implemented as phase gates and the
 single one which involves rotations is implemented via a single Raman pulse. Thus these 
 quantum maps maybe experimentally implementable within these architectures in the immediate
 future. For further details and references we refer the reader to \cite{Milburn}.
  The tilted field changes the character of the dynamics, the Jordan-Wigner transformation
 does not reduce the problem to one of noninteracting fermions and there are features 
 of quantum nonintegrability. This has been studied to some extent in the
  works of Prosen \cite{Prosen}, where he
 has shown different parameter regimes where there is non-ergodic to fully ergodic and mixing
 dynamics in the thermodynamic limit. This model with the tilted field is then one of substantial
 richness which deserves to be further studied in itself. We will use it as a simple and realizable
 model to study the entanglement issues that were discussed in the introduction. It is also worthwhile to
 mention that time evolution can be done with fast numerical algorithms, with a speed up factor 
 of the order of $2^L/L$ to evolve a state one time step, exactly as the fast Hadamard or the 
 fast Fourier transform.
 
  \section{Entanglement in the Integrable cases}

   \subsection{Zero field} 
The simplest nontrivial special case of the models in this paper is an extention of what we discussed
 in the introduction to many qubits. Thus we first discuss the set of states:
\beq
|\psi_L(t)\kt = U_{xx}(J_x)^t |1\kt^{\otimes L}.
\eeq
Here for simplicity we have taken the state with all spins up rather than down. This set 
of states has been discussed earlier \cite{Cluster} and when $t/J_x=\pi, 3 \pi, 5\pi ,\ldots$ the 
states are interesting examples of seemingly highly entangled states. We say 
``seemingly'' as it is not clear that all measures of multipartite entanglement 
will be large for these states, for instance we show below that the $n$-tangle vanishes
for these states, when the number of qubits is larger than $3$. For  
the case of three spins, such states are local unitarily equivalent to the GHZ state. 
Expressing the initial state in the $S_x$ basis we can easily time evolve and 
converting back to the standard $S_z$ basis we arrive at: 
\beq
|\psi_L(t)\kt = \f{1}{2^{L/2}}\sum_{a_k=\{0,1\}} \exp\left(-\f{i J 
t}{4}\sum_{k=1}^{L_0}(2a_k-1)(2a_{k+1}-1)\right)
\bigotimes_{k=1}^{L }\left(\df{|1\kt +(-1)^{a_k}|0\kt}{\sqrt{2}}\right)
\label{clustereqn} 
 \eeq
The states $|\psi_L(\pi/J_x)\kt$ are of special interest. For instance
for $L=2$ we have seen in the introduction that this is essentially one
of the maximally entangled Bell states. Also up to an overall phase
\beq
|\psi_3(\pi/J_x)\kt = \f{1}{2}\left(|111\kt -|100\kt -|010\kt -|001\kt\right),
\eeq 
which after a local phase change $|0\kt \rarrow \sqrt{-1}|0\kt$, $|1\kt \rarrow |1\kt$,
and a $\pi/4$ rotation (Hadamard transform) on each spin becomes the GHZ state
$(|000\kt + |111\kt)/\sqrt{2}$. Similarly up to overall phases we get:
\beqa
|\psi_4(\pi/J_x)\kt = \f{1}{2}\left(|0000\kt -|1111\kt -|1010\kt -|0101\kt\right).&&\\
|\psi_5(\pi/J_x)\kt= \f{1}{4}(|11111\kt  - \hat{\pi} \,(|11100\kt +|10101\kt +|10000\kt)).&&
\label{clusteregs}
\eeqa
 The operation $\hat{\pi}$ on the  states stand for all the five cyclic permutations of this
 one. For these states we have assumed periodic boundary conditions on the spins with $L_0=L$.
 We see however that open chains also give rise to similarly entangled states. It has been
 established earlier that these states with $L>3$ are {\it not} locally convertible to 
 generalized GHZ states by means of LOCC. In some sense that has been termed persistence,
   these states possess higher entanglement content than these N-GHZ states or cat states which 
   are $|0\kt^{\otimes N} +|1\kt ^{\otimes N}$. Persistence is the
   minimum number of local measurements that render the state completely disentangled for all
   possible outcomes \cite{Cluster}.
  In terms of a multipartite generalization of the Schmidt numbers, these states seem to again
   have larger entanglement than the GHZ.

 To calculate the $Q$ measure we find the single qubit reduced density matrix $\rho_k$
  which is 
  \beq
  \rho_k=\left(
  \begin{array}{cc}
  \df{1}{2}-\br S^z_k \kt&\br S^{+}_k \kt\\
  \br S^{-}_k\kt& \df{1}{2}+\br S^z_k \kt
  \end{array}\right)
  \label{singlerdm}
  \eeq
  where the first element is $\br0|\rho_k|0\kt $ etc., and the angular brackets are expectation
  values corresponding to the full pure state $\psi$ we are interested in. The purity is easily expressed
  in terms of these expectation values from which we get the entanglement measure as
  \beq 
  Q(\psi)=  1-\df{4}{L} \sum_{k=1}^{L}\left( \br S^z_k \kt ^2 + |\br S^{+}_k \kt| ^2\right)
  =1-\df{4}{L} \sum_{k=1}^{L}\left( \br S^x_k \kt ^2 + \br S^y_k \kt ^2
  + \br S^z_k \kt ^2 \right)
  \eeq
For the states under consideration $|\psi_L(t)\kt$ we may explicitly calculate these to get
\beq
\br S^z_k \kt= \df{1}{2}\cos^2(J_x\,t/2),\;\;\br S^+_k \kt=0,
\eeq
and hence
\beq
Q(\psi_L)= 1-\cos^4(J_x\,t/2). 
\eeq
Thus this measure of entanglement for this class of states is independent of the
 length of the chain $L$, and periodically reaches a maximum at
  $t=\pi/J_x,\, 3 \pi/J_x, \ldots$, as indicated earlier, and this maximum is the highest
  possible. At $t=0, 2 \pi/J_x, \ldots$ the state is completely unentangled and therefore
  in this simple time evolution we have large entangling and disentangling oscillations. The
  periodic boundary condition can be replaced by an open chain, in which case the entanglement
  content as measured by $Q$ is
  \beq
  Q_{open}= 1-\cos^4(J_x\,t/2) -\df{1}{2L}\sin^2(J_x \,t),
  \eeq
 implying again maximal entanglement at times that are odd multiples of $\pi/J_x$. Notice that
 for open chains there is marginal dependence of $Q$ on the number of spins, and for $L=2$
 this simplifies to $\sin^2(J_x\,t/2)$, which we have already derived as the 2-tangle
 for this state in the introduction.
 
 If for these states there is high entanglement content as measured by $Q$, the two-spin
 correlations as measured by the concurrence is of interest. For $L=2$ the (square of the)
 concurrence coincides with $Q$, but for higher number of spins, we find that while nearest
 neighbour concurrences persist and oscillate in time, all other concurrences are perpetually
 and strictly zero. Also the times at which the nearest neighbour concurrence vanish are
 periods when the multipartite entanglement content as measure by $Q$ is maximized, indicating that
 two-body correlations are being distributed more globally.
   To calculate the concurrence between any two spins, at say positions $i$ and $j$, of the chain, we need the two-spin
 reduced density matrix which is
 \beq
 \br ab|\rho_{ij}|cd\kt =\sum_{s_k\in\{0,1\}} \br s_1s_2 \ldots a \ldots b \ldots s_L|\psi\kt
  \br \psi |s_1s_2\ldots c \ldots d \ldots s_L\kt, 
  \eeq   
 where $a$ and $c$ are fixed states at position $i$ ($0$ or $1$) and similarly $b$ and $d$ are
 at position $j$. This matrix can also be written in terms of spin-expectation values as
 \beqa
 \br 00|\rho_{ij}|00\kt = \br(\df{1}{2}-S_i^z) (\df{1}{2}-S_j^z)\kt, \,\br 00|\rho_{ij}|01\kt
  =   \br(\df{1}{2}-S_i^z) S^{+}_j\kt,\, \br 00|\rho_{ij}|10\kt = 
 \br S^{+}_i(\df{1}{2}-S_j^z)\kt, \nonumber \\ \br 00|\rho_{ij}|11\kt
  =\br S^{+}_i S^{+}_j\kt,\,
   \br 01|\rho_{ij}|01\kt = \br(\df{1}{2}-S_i^z) (\df{1}{2}+S_j^z)\kt, \,
   \br 01|\rho_{ij}|10\kt =\br S^{+}_i S^{-}_j\kt,\nonumber \\ \br 01|\rho_{ij}|11\kt = 
 \br S^{+}_i(\df{1}{2}+S_j^z)\kt,\, \br 10|\rho_{ij}|10\kt = \br(\df{1}{2}+S_i^z)
  (\df{1}{2}-S_j^z)\kt, \nonumber \\
\br 10|\rho_{ij}|11\kt
  =   \br(\df{1}{2}+S_i^z) S^{+}_j\kt,\,
  \br 11|\rho_{ij}|11\kt = \br(\df{1}{2}+S_i^z) (\df{1}{2}+S_j^z)\kt.
 \label{doublerdm}
 \eeqa
 
 The rest of the matrix elements follow from Hermiticity of the density matrix.
 For the class of states given by $|\psi_L(t)\kt$ we can calculate these expectation values
 in a straightforward manner, exploiting the translational symmetry of the states.
 We get that if $j\ne i \pm 1$ that the density matrix is diagonal, in fact
 \beq
 \rho_{ij}=\rho_i \otimes \rho_j, \;\; j \ne i \pm 1. 
 \eeq
 Here $\rho_i$ and $\rho_j$ are the single spin density matrices as given
  in Eq.~(\ref{singlerdm}). 
Thus there is no concurrence between spins that are not nearest neighbours. 
For the case when $j=i \pm 1$ we get that 
\beq
\br 00|\rho_{i,i \pm 1}|00\kt= \br 01|\rho_{i,i\pm1}|01\kt=
\br 10|\rho_{i,i \pm 1}|10\kt= \df{1}{4}\sin^2(J_x\,t/2),\, \br 11|\rho_{i,i\pm 1}|11\kt =
1-\df{3}{4}\sin^2(J_x\,t/2).
\eeq
The only nonzero off-diagonal matrix element is 
\beq
\br 00|\rho_{i,i \pm 1}|11\kt=\df{-i}{4}\sin(J_x\,t)
\eeq

 For density matrices such as we have, with
   all vanishing off-diagonal elements except the corner ones, it is easy to find the 
   concurrence in terms of the matrix elements of the density matrix itself. The positive
   square-roots of the eigenvalues of the matrix $\rho_{i,i \pm 1}\, \tilde{\rho}_{i,i\pm1}$,
   arranged in nonincreasing order are $|b|+\sqrt{a(1-3a)},\,a,\,a,\,-|b|+\sqrt{a(1-3a)} $, where 
   $a= \sin^2(J_x\,t/2)/4$ and $b=|\sin(J_x\,t)/4|$.
   Thus in this case we get,
  \beq 
  C_{i,i\pm1}(t)=\mbox{max}\left(0, \, \f{1}{2}\left( \left| \sin(J_x\, t)\right|-\sin^2(J_x\, t/2)
   \right) \right).
   \eeq
   Thus we can explicitly calculate the concurrences at all times, as we see that times at which
   $Q$ is a maximum namely at $t=\pi/J_x, 2\pi/J_x , \ldots$, {\it all} the concurrences
   vanish, including nearest neighbour ones. In fact there is a period of time around 
   when  $Q$ reaches its maximum that there is no two-body entanglement at all.
   We get that $C_{i,i\pm1}=0$ if $|\tan(J_x t/2)|>2$ or if
    $t=2\pi k/J_x$, $k=0, \pm 1, \pm 2, \ldots$.

\begin{figure}
\includegraphics[width=2.5in,angle=-90]{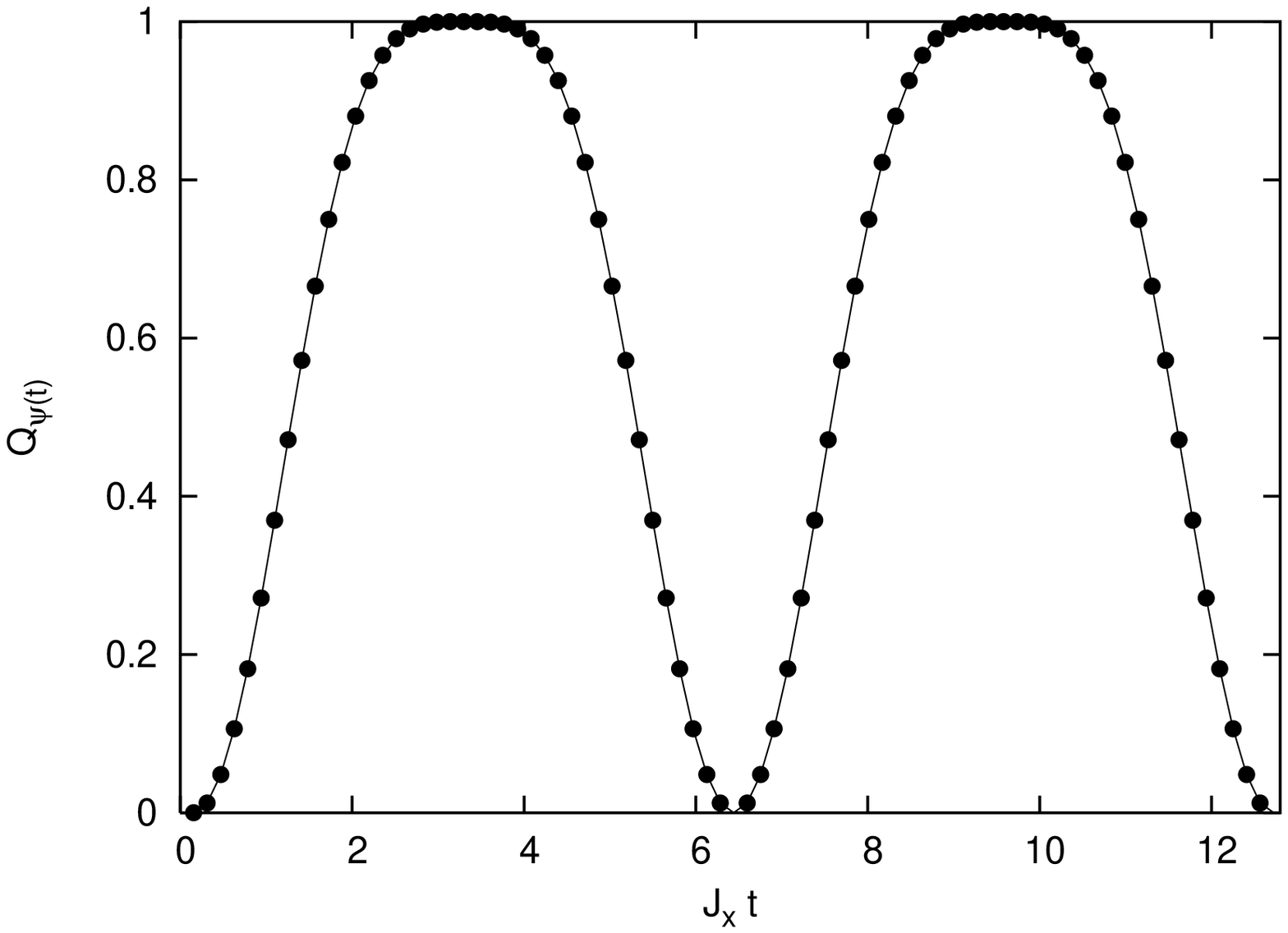}
\includegraphics[width=2.5in,angle=-90]{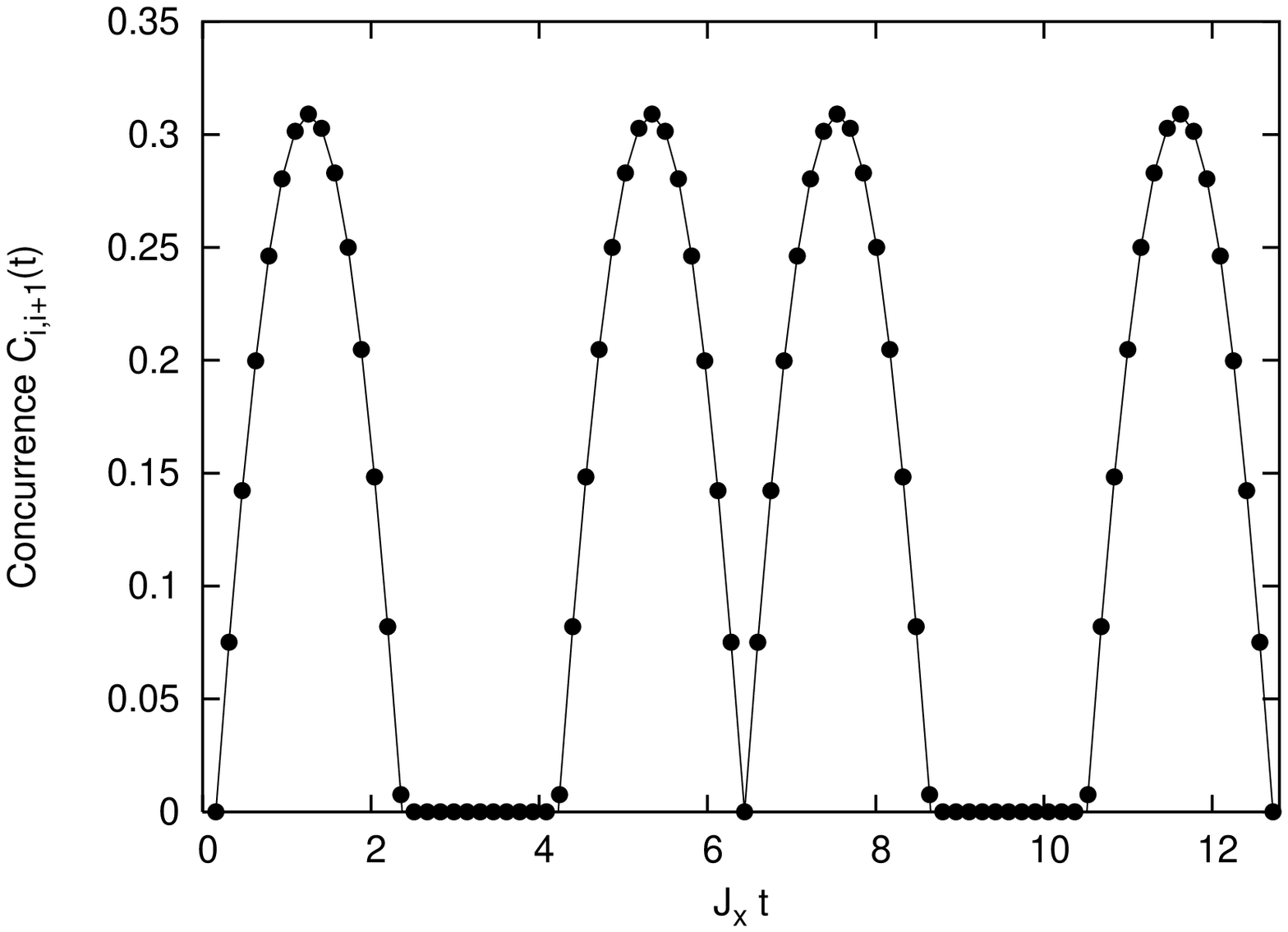}
\caption{The Meyer and Wallach measure of entanglement $Q$ and the nearest neighbour 
concurrence for the state $|\psi_4(t)\kt$ as 
functions of (scaled) time. Plotted are the numerical (points) and the formula (solid line). Periodic
boundary conditions were used.} 
\label{SxSxQCon}
\end{figure}

\begin{figure}
\includegraphics[height=4in,angle=-90]{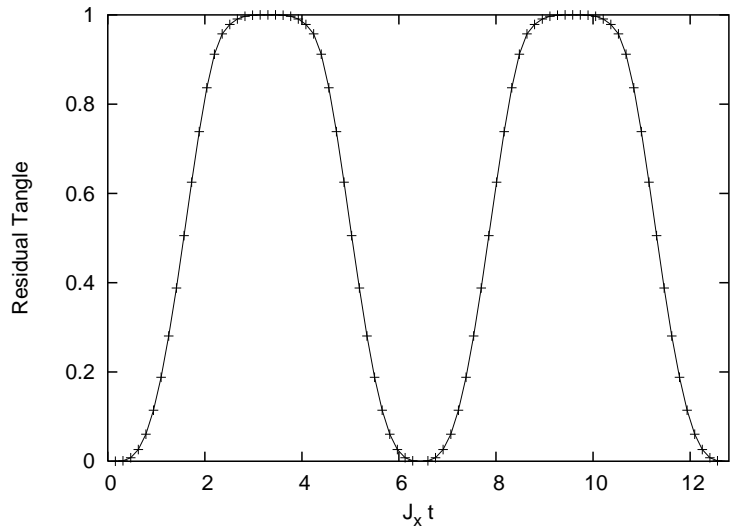}
\caption{The residual tangle for the state $|\psi_4(t)\kt$ as a
function of scaled time. Plotted 
are the numerical (points) and the formula (solid line). Periodic boundary conditions were
used.} 
\label{SxSxT}
\end{figure}

In Fig.~(\ref{SxSxQCon}) is shown the entanglement measure $Q$ and the nearest 
neighbour concurrence as a function of time. This figure is independent of the number of 
spins in the chain, as long as $L>2$. The concurrence is dominated by other types 
of entanglement. It has been conjectured that \cite{CoffValWoot} 
\beq
\tau_i \, - \, \sum_{j\ne i} C_{i,j}^2 \, \ge \, 0.
\eeq 
As we have shown earlier the average of $\tau_i$ is nothing but the entanglement measure $Q$,
and from translational invariance of the states under discussion, this is also equal to
any $\tau_i$. For this class of states the inequality is easily seen to be rigorously true.
 This difference is interpreted as the generalization of the residual tangle, 
 entanglement not present in the form of two-body correlations.
In Fig.~(\ref{SxSxT}) is plotted this residual tangle which is dominated by the tangle
 of individual spins with the others, and not by the concurrence. 

Although both $Q$ and the residual tangle are maximum for states such as $|\psi_4(\pi/J_x)\kt$
 the $n$-tangle measure, as previously stated, {\em vanishes} for these
 states. We may calculate 
explicitly this measure for the states in Eq.(\ref{clustereqn}), and we find that 
\beq
\tau_N(\psi_L) =\df{1}{2^{L-2}} \sin^{L}(J_x \,t).
\eeq
Thus the $n$-tangle decreases exponentially with the number of qubits for the cluster 
state, it seems to be a rare entanglement feature, and in particular 
for the states at $t/J_x=\pi, 2\pi , \ldots$, such as those in Eq.(\ref{clusteregs})
the $n$-tangle vanishes. We note in passing that one class of states for which both the 
$n$-tangle and the $Q$ measures are high are easily obtained from the states
discussed here so far by symmetrizing with respect to the ``up'' and ``down'' spins.
Thus we consider initial states that are N-GHZ states, with the dynamics of 
nearest neighbour coupling. 
\beq
|\phi_L(t)\kt = U^t _{xx}\, \df{1}{\sqrt{2}}\left(|0\kt^{\otimes L} + |1 \kt^{\otimes L} \right)=
\df{1}{\sqrt{2}}( 1 + \otimes_{k=1}^{L} \sigma^x)  |\psi_L(t)\kt.
\eeq
The last equality follows since the time evolution commutes with the operator $\sigma^x$
 that flips spins in the standard basis. For these states $\br S^z_k \kt = \br S^{+}_k\kt =0$ for all $k$, implying that the single spin reduced
density matrix is maximally mixed, and the measure $Q$ is unity for all
time $t$. The $n$-tangle though changes from the maximal value of unity at zero time (the 
N-GHZ states) and oscillates with exact returns to unity at multiples of $\pi/J_x$. 
\beq
\tau_N(\phi_L)= \left| \cos^{L/2} (J_x t/2) \, +\ i^{L/2} \sin^{L/2}(J_x t/2)\right|^4.
\eeq
Thus for this class of symmetrized cluster states, the $n$-tangle does not decrease
exponentially with the number qubits and can have the maximal value at nonzero times.
We remind the reader that this measure requires that the number of qubits $L$ be even.

For $L=3$ the state $\phi$ is 
\beq
|\phi_3(\tau)\kt = \df{1}{2\sqrt{2}}\left( \left( e^{i\tau/2}+ \cos(\tau/2) \right)
 (|000\kt +|111\kt)\, -\, i \, \sin(\tau/2) \left(\hat{\pi}(001+110)\right) \right),
 \eeq 
where we have written the scaled time $\tau=J_x t$, that may be simply viewed as a real
parameter. For three qubits the residual tangle and 
provides a global entanglement measure \cite{CoffValWoot}.
 For the state $|\phi_3(\tau)\kt$, the one and two
spin reduced density matrices are simply $\rho_1=I_2/2$, where $I_2$ is the two dimensional
identity operator and
\beq
\rho_{12}= \df{1}{4}\left(\begin{array}{cccc}
1+\cos^2(\tau)& i \sin(2\tau)/2 & i \sin(2 \tau)/2 & -\sin^2(\tau)\\
&\sin^2(\tau)&\sin^2(\tau)& -i \sin(2 \tau)/2\\
& & \sin^2(\tau)&-i \sin(2\tau)/2\\
& & & 1+\cos^2(\tau)
\end{array}\right).
\eeq 
The other matrix elements of $\rho_{12}$ follow from Hermiticity of this matrix.
Due to translational invariance these are the only relevant operators. The spectrum of 
$\rho_{12}$ is $\{0,0,1/2,1/2\}$, independent of the parameter $\tau$ and the spectrum of 
$\rho_{12}\tilde{\rho_{12}}$ is similarly $\{0,0,1/4,1/4\}$. Thus the concurrence vanishes
between any two qubits for all values of the parameter (time) $\tau$. The tangle between one
qubit and the other two is $\tau_1=4 \det{\rho_1}=1$, thus the residual tangle is
$\tau_1 -C_{1,2}^2 -C_{1,3}^2=1$. Thus we have a continuous one parameter family of states,
of which the GHZ state is a special case, that have maximal entanglement, as measured
 by both $Q$ and the residual
tangle. Note that in the case of $3$ qubits the residual tangle is also maximized for all time, a 
feature that generalizes to higher number of qubits, while the $n$-tangle oscillates as
 indicated above.

 For $L=4$ the state is 
 \beq
 |\phi_4(\tau)\kt = \f{1}{\sqrt{2}}\cos^2(\tau/2) (|0000\kt +|1111\kt)-\f{i}{2\sqrt{2}}\sin(\tau)
 \hat{\pi}(|1100\kt) -\f{1}{\sqrt{2}}\sin^2(\tau/2)(|1010\kt +|0101\kt).
 \eeq
 While $Q(\phi_4)=1$ for all $\tau$, the $n$-tangle is maximized for $\tau=\pi$, in which case
 the state becomes proportional to $|1010\kt +|0101\kt$, which is local unitarily equivalent to the
  4-GHZ state, by say flipping the first and third spins. However for larger number of qubits,
   the state that maximized the $n$-tangle
  is apparently not the N-GHZ state. For instance for $L=6$, and $8$  we get
  \beqa
  |\phi_6(\pi)\kt &=& \f{1}{4 \sqrt{2}}\left(000000 +\hat{\pi}(101000+100100-110000)+ 1\leftrightarrow0\right).\\
  |\phi_8(\pi)\kt &=& \f{1}{4\sqrt{2}}\left(00000000+\hat{\pi}(00010001-01100110+10101010-00001111+01000100)
  + 1\leftrightarrow 0\right).
  \eeqa
  There are a total of $32$ terms in each state and we have temporarily dispensed with the ket notation.

\subsection{Transverse field}

We now turn on an external field in the transverse direction. This model, the kicked transverse 
Ising model, has been studied recently as noted above
 and is also an integrable case \cite{Prosen,Milburn}, and the Jordan~-~Wigner transformation
  can be used to diagonalize it. In this case we have 
 \beq
 |\psi_L(t)\kt = \left(U_{xx}(J_x) U_{x,z}(B,\pi/2)\right)^t |\psi_L(0)\kt
 \eeq
where $t$ is an integer time, the number of kicks. We now proceed to diagonalize the
operator, indicating the key steps. It maybe noted that unlike the treatment in \cite{Milburn}
we do not assume the thermodynamic limit, and in this sense the way we solve this problem
is also new, though the technique is the same as that for the usual Ising model in a 
transverse field.
 
In the kicked transverse Ising spin chain treated here, the Ising interaction
 is in $x$-direction and the magnetic field is switched on at integer times along the $z$-direction.
The first step is to replace the spin variables by Jordan-Wigner fermions
through a nonlocal transformation \cite{JordanWigner}:
\begin{equation}
S_l^+=\exp \left(i\sum_{n=1}^{l-1}c_n^{\dag}c_n \right) \, c_l^{\dag},\; S_l^z=c_l^{\dag}c_l-
\f{1}{2}.
\end{equation}
The operators $c_l$ and $c_l^{\dag}$ obey the usual fermion anticommutation rules.
The interaction term in $U_{xx}$ reduces to a combination of nearest-neighbour
fermion hopping, pair-fermion annihilation and creation terms on a lattice,
\begin{equation}
U_{xx}=\exp\left( -\df{i J_x}{4} \left( \sum_{l=1}^{L-1} (c_l^{\dag}-c_l)(c_{l+1}^
{\dag}+ c_{l+1}) - (-1)^{N_F} (c_L^{\dag}-c_L)(c_1^{\dag}+c_1) \right) \right)
\end{equation}
where $N_F=\sum_{i=1}^{L} c_i^\dag c_i$ is the total number of fermions.
 The last term is due to the periodic boundary condition.
The magnetic field term in $U_{x,z}(B,\pi/2)$ becomes a chemical potential term for
the total number of fermions. The eigenstates of $U$ will
have a definite even or odd fermion number, since $N_f$ commutes with $U$,
 and we can find the eigenstates in the two sectors separately. 

Now, the second step is to Fourier transform through, 
\begin{equation}
c_q=\df{\exp\left( i \pi/ 4\right)}{ \sqrt{L}}\sum_{l=1}^{L} \exp\left(-iql\right) c_l,
\end{equation}
where the allowed allowed values for $q$ are (taking $L$ to be even)
\begin{eqnarray}
q=\pm \df{\pi}{L},\pm \df{3\pi}{L}, \ldots, \pm\df{(L-1)\pi}{L}~~~~~~~~~~~~~~N_F 
~~{\rm even},\\
q=0,\pm \df{2 \pi}{L},\pm \df{4\pi}{L}, \ldots, \pm\df{(L-2)\pi}{L},\pi~~~~~~~~~N_F 
~~{\rm odd.}
\end{eqnarray}
The lattice momentum $q$ labels the momentum creation and annihilation operators that 
also obey the fermion anticommutation rules. The unitary operator $U$ 
has a direct product structure in terms of these fermion variables:
\begin{eqnarray}
U&= {\rm e}^{-i\f{BL}{2}} \prod_{q>0} V_q~~~~~~~~~~~~~~~~N_F~~{\rm even}, \\
&={\rm e}^{-i\f{BL}{2}}V_0V_{\pi} \prod_{q>0} V_q~~~~~~~~N_F~~{\rm odd}
\end{eqnarray}
where 
\begin{equation}
V_q= \exp\left(-i \df{J_x}{2}\left[ \cos(q) 
( c_q^{\dag} c_q+c_{-q}^{\dag}c_{-q})
+\sin(q) (c_q c_{-q} +c_{-q}^{\dag}c_q^{\dag})\right] \right)~~
\exp\left(-iB ( c_q^{\dag} c_q + c_{-q}^{\dag} c_{-q})\right),
\end{equation}
and
\begin{equation}
V_0=\exp\left(-i(B+\f{J_x}{2})c_0^{\dag}c_0\right),~~~V_{\pi}=
\exp\left(-i(B-\f{J_x}{2})c_{\pi}^{\dag}c_{\pi}\right).
\end{equation}
The eigenstates of $U$ are direct products of eigenstates of $V_q$. The
operators $V_0$ and $V_{\pi}$ are diagonal in the number basis states. For
$V_q$, the four basis states are $|0 \kt,|\pm q \kt=c_{\pm q}^{\dag}|0 \kt,|-qq>=c_{-q}^
{\dag}c_{q}^{\dag}|0 \kt$.
 The eigenstates of $V_q$, for $q\ne 0,\pi$ are given
by
\begin{equation}
V_q|\pm q \kt= {\rm e}^{-i(\f{J_x}{2}+B)} |\pm q \kt,~~~V_q|\pm \kt=  
{\rm e}^{-i(\f{J_x}{2}+B)} {\rm e}^{\pm i \theta_q} |\pm \kt.
\end{equation}
Here the eigenstates $|\pm\kt $ are given by $|\pm \kt\,  \equiv \, a_{\pm}(q)|0 \kt \, +\, 
b_{\pm}(q)|-qq \kt$. Using $\cos(\theta_q)=\cos(B)\cos(J_x/2)-\cos(q) \sin(B)
\sin(J_x/2)$, we have
\beq
a_{\pm}(q)^{-1}\, =\, \sqrt{ 1+\left(\df{\cos(J_x/2)-\cos(\theta_q\pm B)}{\sin{q} \sin{B}
\sin(J_x/2)}\right)^2},
\eeq 
\beq
b_{\pm}(q)=a_{\pm}(q)\df{\pm\sin(\theta_q) + \cos(J_x/2)
\sin{B}-\cos{q}\cos{B}\sin(J_x/2)}{ \sin(q)\sin(J_x/2)} e^{
-i2B}.
\eeq

\begin{figure}
\includegraphics[height=4in,angle=-90]{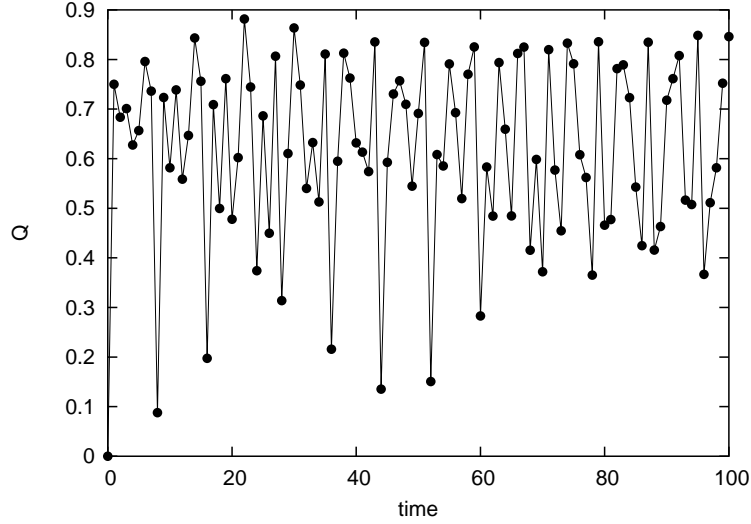}
\caption{The measure $Q$ for the kicked transverse Ising interaction, when the initial
state is the vacuum state and $L=10$, and the parameters are $J_x=\pi/2$, $B=\pi/3$. Shown are   
the results of  the numerical calculations (points) and using the formula (solid line). Periodic boundary conditions are
assumed.} 
\label{TIM1}
\end{figure}    
\begin{figure*}
\includegraphics[height=3.5in,angle=0]{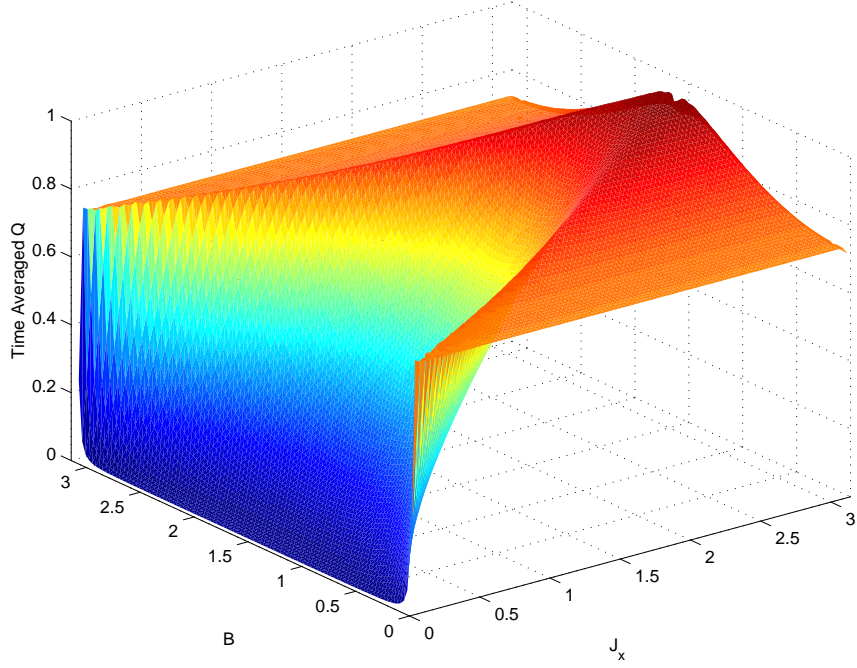}
\caption{(Color Online) The time averaged $Q$ as a function of system parameters for the kicked transverse
Ising model. $L=20$ in this case, and the averaging is done over a thousand kicks, by which time the
average is stationary.}
\label{tnisi2}
\end{figure*}      

This then completely solves the kicked transverse Ising model.
Let us consider an initial state with $m$ (even) fermions $|\psi(t=0) \kt=|l_1,
l_2...l_m \kt$
where $l_i$ denote the sites occupied by fermions (corresponding to $S_{l_i}^z=
1/2$ in terms of the original spin variables). The off-diagonal matrix element
of $\rho_l$ through time evolution with $U$ is
\begin{equation}
\br S^{+}_l(t) \kt \equiv = \br \psi(t)|e^{i\pi\sum c_n^{\dag}c_n} c_l^{\dag}|\psi(t) \kt=0,
\end{equation}
as the time evolution mixes only states with even number of fermions.
The diagonal matrix elements of $\rho_l$ depend on $\br S_l^z \kt \equiv \br \psi(t)|
c_l^{\dag}c_l|\psi(t) \kt -1/2$. This can be calculated from the time-evolved
operator,
\begin{equation}
c_q(t)=V_q^{\dag t} \, c_q \,  V_q^t= \zeta_q \, c_q \,  - \, \mbox{sgn}(q) \, \eta_q \,c_{-q}^{\dag},
\end{equation}
where the expansion coefficients are given as
\begin{eqnarray}
\zeta_q &=& |a_+(q)|^2 {\rm e}^{-it\theta_q}+ |a_-(q)|^2 {\rm e}^{it\theta_q},\\
\eta_q &=& a_+(q)^* b_+(q) {\rm e}^{-it\theta_q}+ a_-(q)^*b_-(q) {\rm e}^{it
\theta_q}.
\end{eqnarray}
The diagonal matrix element can be expressed in terms of the Fourier transforms
of the above functions, after some manipulations, we have
\begin{equation}
\br S_l^z(t)\kt=-\df{1}{2} + \df{1}{L} \, \sum_{q} |\eta_q|^2  \, + \, \sum_{i=1}^m \, |\zeta(
l-l_i)|^2 \, - \, |\eta(l-l_i)|^2.
\end{equation}
In the above we used two more auxiliary functions defined by
\begin{eqnarray}
\eta(l)&=&\df{2}{L} \sum_{q>0} \eta_q \cos{(ql)},\\
\zeta(l)&=&\df{2}{L} \sum_{q>0} \zeta_q \cos{(ql)}.
\end{eqnarray}

In particular for the initial unentangled state $|\psi_L(0)\kt=|0\kt^{\otimes L}$,
as a special case we can calculate $\br S^z_l(t) \kt$ at any site using the above.
\beq
\br S^z_l(t) \kt = \br \psi_L(0)|S^z(t)|\psi_L(0)\kt = \df{1}{L} \sum_{q}|\eta_q|^2 -1/2.
\eeq
Here the $q$ summation extends to both positive and negative allowed values.
Hence using translational symmetry the entanglement measure $Q$ is given in this case 
by  
\beq
Q(\psi_L(t)) = 4 \, x\, (1-x), \; x= \df{1}{L} \sum_{q}|\eta_q|^2 = \df{4}{L} \sum_{q}
 |a_+(q) a_-(q) \sin(\theta_q\, t)|^2
 \label{TIMQ}
\eeq

As illustrated in the example (Fig.~(\ref{TIM1})) the oscillations of $Q$ are now much more
complicated. The advantage of having an easily computable formula such as Eq.~(\ref{TIMQ})
 is that we can study the
entanglement measures as a function of the interaction strength and transverse magnetic
fields more comprehensively. In order to do that we time average $Q$ over sufficiently long
scales and plot this as a function of $J_x$ and $B$ in Fig.~(\ref{tnisi2}). This figure shows
some interesting features, especially the large $Q$ parts which correspond to $J_x=\pi$. Note that 
both the lines $B=0$ and $J_x=0$ have been discussed previously, the latter case turns 
off the interaction and produces no entanglement, while the former is the zero field 
case for which the cluster states were realized.

The case when $J_x=\pi$, $B=\pi/2$ simplifies considerably, as in this case $a_{\pm}(q)=1/\sqrt{2}$
and $\theta_q=\pi - q$. Thus $Q$ can be calculated more explicitly and results in 
\beq
Q(t)=\left\{ \begin{array}{cl} 1& \mbox{if $t \ne kL/2$}\\0& \mbox{if $t = kL/2$} \end{array}
\right.\; \; k=0,1,2, \ldots.
\eeq 
Thus either the state is maximally entangled by the measure $Q$ or is not at all entangled.
As in the zero field case, if the initial state is an N-GHZ state itself, according to the
$Q$ measure it remains maximally entangled, as in this case also $\br S^z_k\kt=\br S^+_k \kt =0$
for all times. As in that case the $n$-tangle measure is now significant, although not maximal
in general. In the case $B=\pi/2,\, J_x=\pi$ both the $Q$ measure and the 
$n$-tangle are unity and represent highly entangled states, which appear to be in the nature of
cluster states discussed previously for the zero field case. Incidentally, this point is also
on the critical line $J_x=2B$ of the (unkicked) transverse Ising model.

\section{Entanglement in the Nonintegrable case}

We now consider the case when the field is tilted in the $x-z$ plane, that is the 
unitary operator is a slight modification of the transverse Ising case:
\beq
 |\psi_L(t)\kt = \left(U_{xx}(J_x) U_{x,z}(B, \theta)\right)^t |\psi_L(0)\kt.
 \eeq
The case when $\theta$ is different from both zero and $\pi/2$, as has been noted earlier,
constitutes a nonintegrable model. The Jordan-Wigner transformation no longer
renders the problem into one of noninteracting fermions. Here we study the influence of this
the entanglement content of the states $|\psi_L(t)\kt$, again when the 
initial state is the ``vacuum'' state. Once more the time $t$ takes integer values.
Since the Jordan-Wigner transformation does not help, much of the results in this
section are done purely numerically, with the help of the fast Hadamard transform.

\begin{figure}
\includegraphics[width=4in,angle=-90]{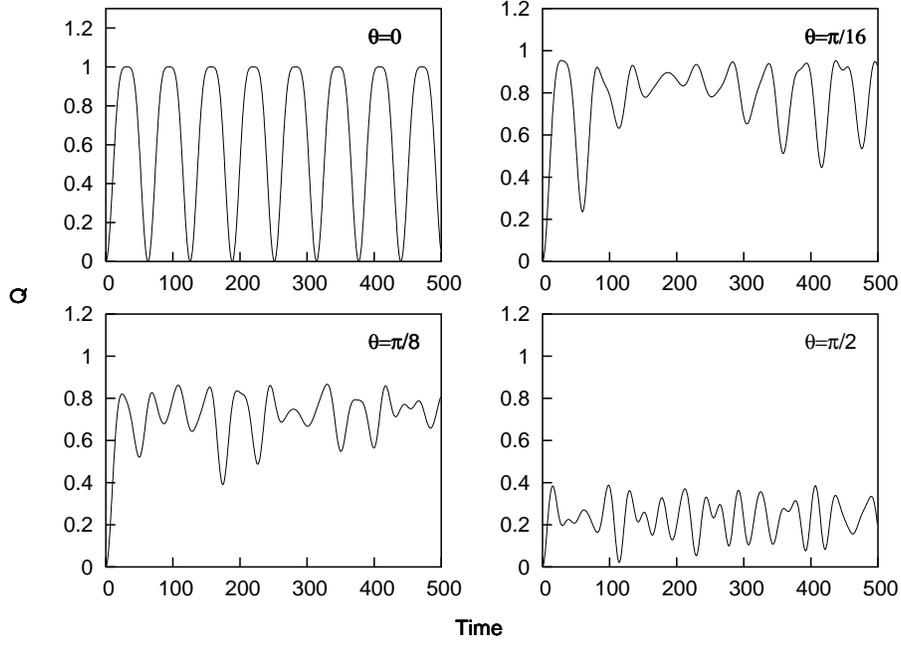}
\caption{The entanglement measure $Q$ as a function of time, for different tilt
angles of the external field. The parameters are  $J_x=0.1$, $B=0.1$, $L=10$.}
\label{tilt1A}
\end{figure}

\begin{figure}
\includegraphics[width=4in,angle=-90]{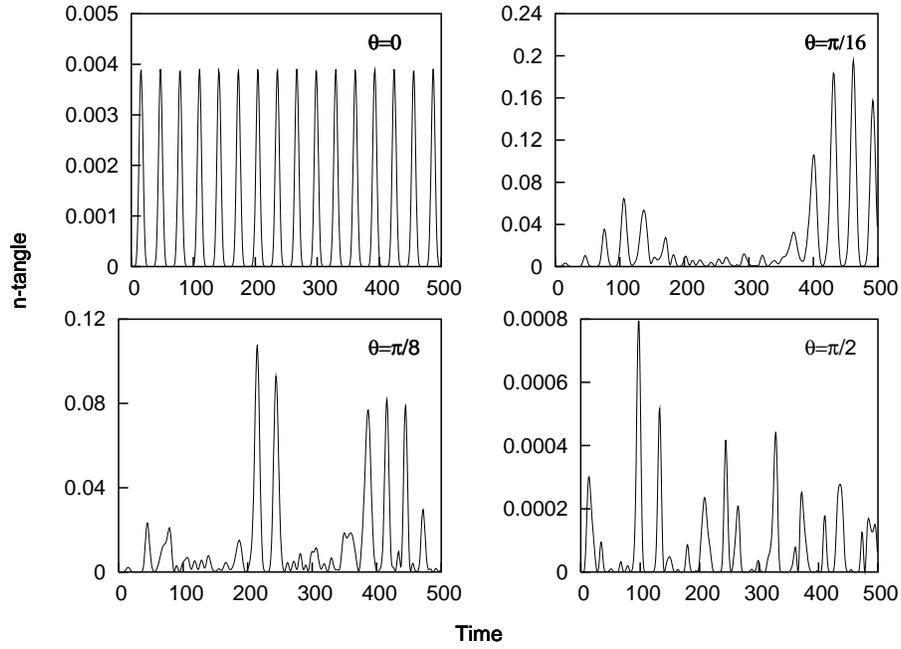}
\caption{The $n$-tangle as a function of time, the parameters and tilt
angles of the external field are the same as in the previous figure.}
\label{tilt1B}
\end{figure}

We start with a given exchange coupling, and strength of the external field, while varying the
angle of tilt of this field from zero to ninety degrees, both these extremes being 
integrable. In Figs.~(\ref{tilt1A},\ref{tilt1B}) we see the result of this for a particular case. We
note that  the $\theta=0$ case is integrable and is essentially the zero field case
we have discussed earlier. In this case the $Q$ measure of entanglement reaches the maximum
value of unity and drops back to zero periodically. With a non-zero tilt angle we see that the
while the maximum drops from unity, the propensity to untangle also decreases considerably,
thereby providing on the average larger entanglement than for the zero-tilt case. Increasing
 the angle of tilt further decreases the typical value of entanglement produced. The $n$-tangle
 measure shows more complicated behaviour, with an intermediate angle producing states that 
 have a large $n$-tangle. 
 
\begin{figure}
\includegraphics[width=4in,angle=-90]{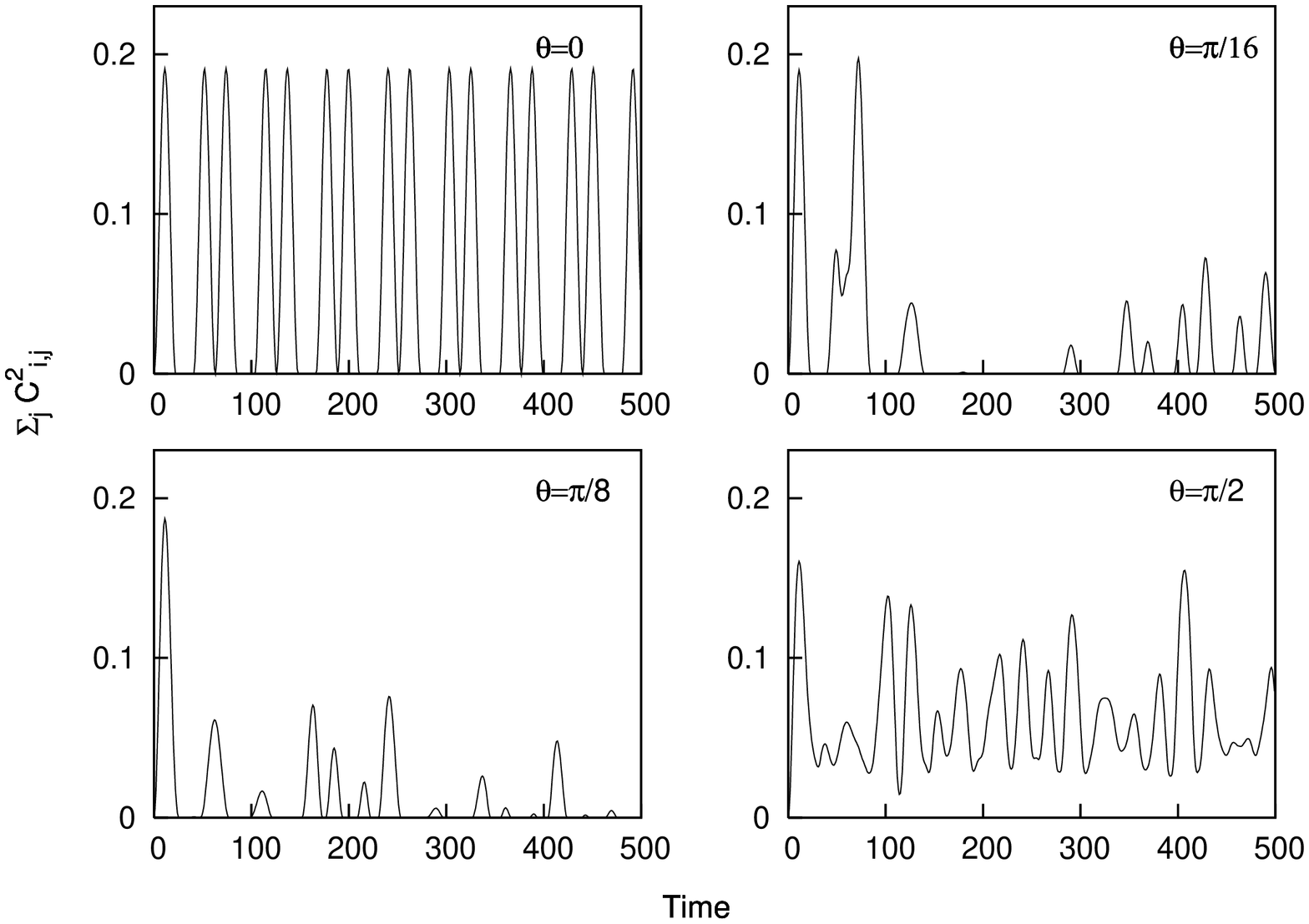}
\caption{The sum of the two-body tangles as a function of time, for various tilt
angles of the external field. The parameters are  $J_x=0.1$, $B=0.1$, $L=10$.}
\label{figtiltconc1}
\end{figure}

The increase in the average multipartite entanglement as measured by $Q$
 is accompanied by decreasing overall two-body correlations, as captured by the
 pairwise concurrence amongst the qubits. This is illustrated in Fig.~(\ref{figtiltconc1})
where we show the sum of the two-tangles (or the square of the concurrences), between a 
given qubit and the rest of them. Due to translational symmetry the sum is independent
 of the focus qubit. This figure shows the rather substantial concurrences that are present
  in the integrable cases (both $\theta=0$ and $\theta=\pi/2$), compared to the
   nonintegrable ones. Thus like the
GHZ state that has no two-body correlations, such as the concurrence, these appear to be
highly entangled states with small concurrences. The entanglement present in the state appears
to be predominantly not of the two-body type. In fact we noted this previously for the cluster
states that when $Q$ was the maximum possible the concurrences identically vanished. 

\begin{figure*}
\includegraphics[width=3in,angle=-90]{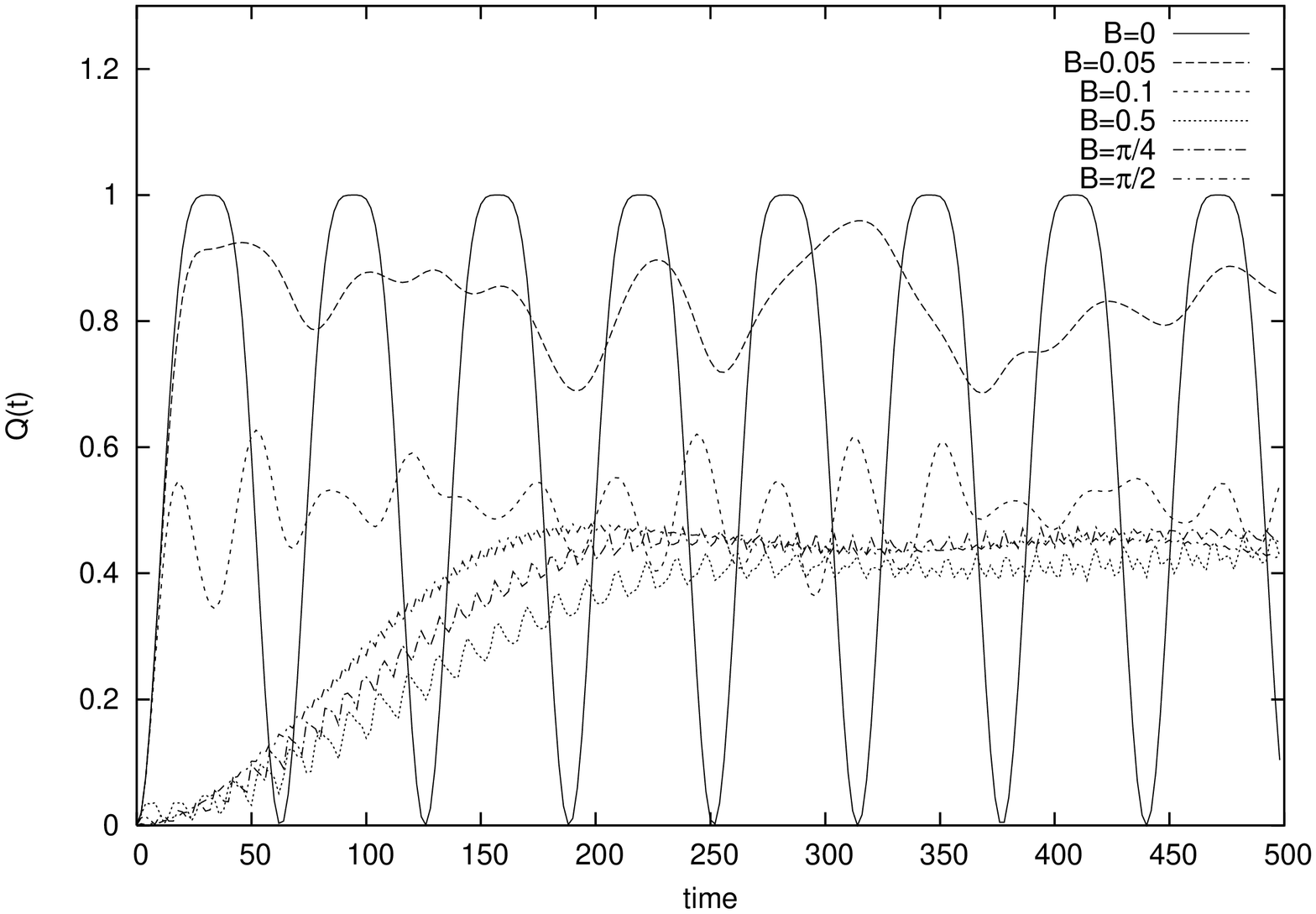}
\caption{The entanglement measure $Q$ as a function of time, for various magnitudes of
the external field. The parameters are  $J_x=0.1$, $\theta=\pi/4$, $L=10$.}
\label{tilt3A}
\end{figure*}

\begin{figure*}
\includegraphics[width=3in,angle=-90]{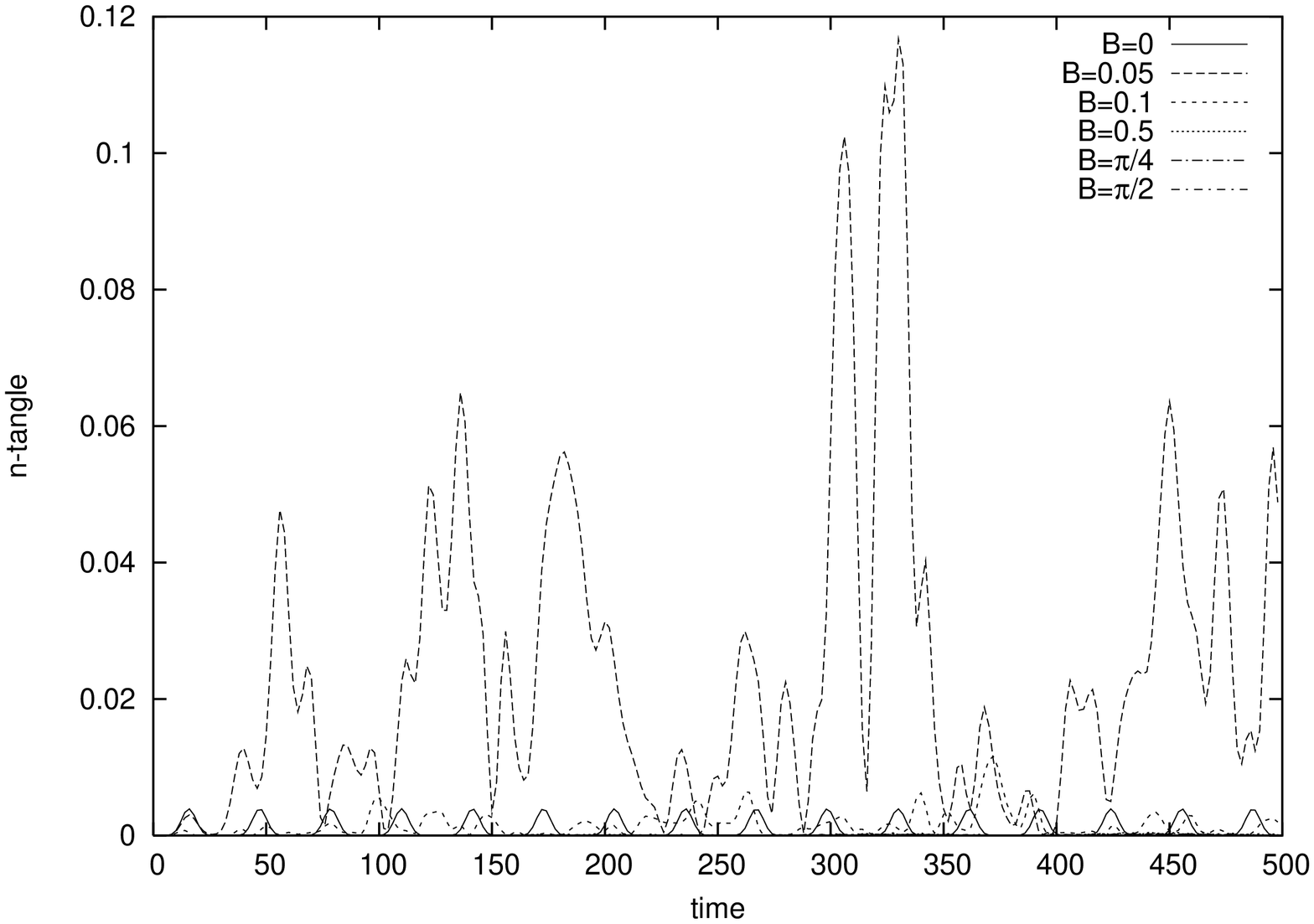}
\caption{The $n$-tangle as a function of time, for various magnitudes of
the external field. The parameters are same as that of the previous figure.}
\label{tilt3B}
\end{figure*}

Thus it appears that both the $Q$ and $n$-tangle measures are sensitive to the 
nonintegrability of the spin chain and from this preliminary data it is plausible
that entanglement is enhanced on the average. We have found this to be the case 
for other values of the parameters, not shown here. We can hold the angle fixed
 and vary the magnitude of the external field. Both Figs.~(\ref{tilt3A},\ref{tilt3B})
  are of this kind. In this case it is seen that small 
 values of the magnetic field are enough to prevent the states from completely disentangling.
 Larger fields also bring down the average along with the fluctuations, till for sufficiently
 large fields the chain seems to reach smoothly an entanglement plateau. The $n$-tangle again
 shows more complicated behaviour, and can be substantially large in comparison with the 
 integrable cases.
  
\begin{figure*}
\includegraphics[width=2in,angle=-90]{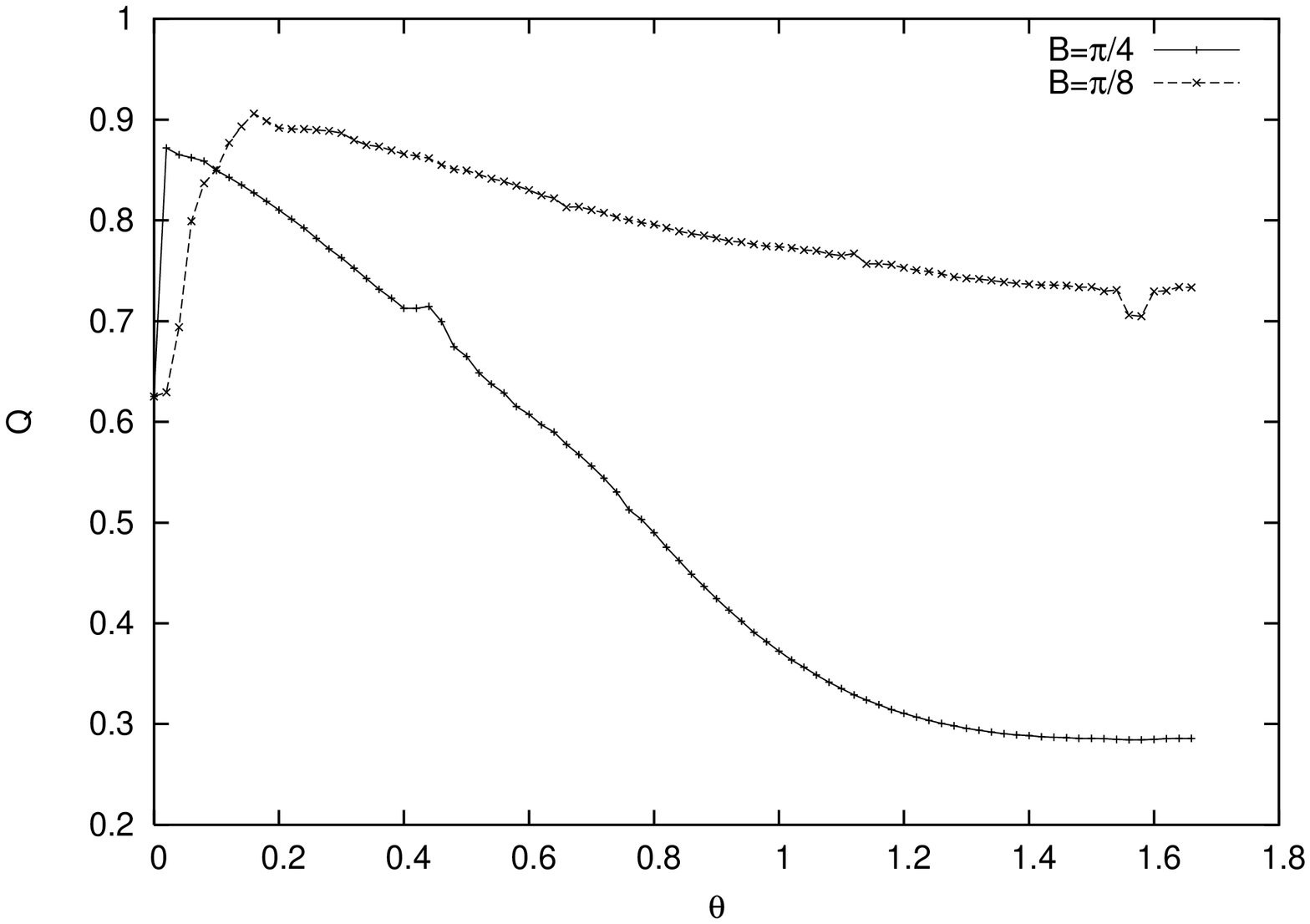}
\includegraphics[width=2in,angle=-90]{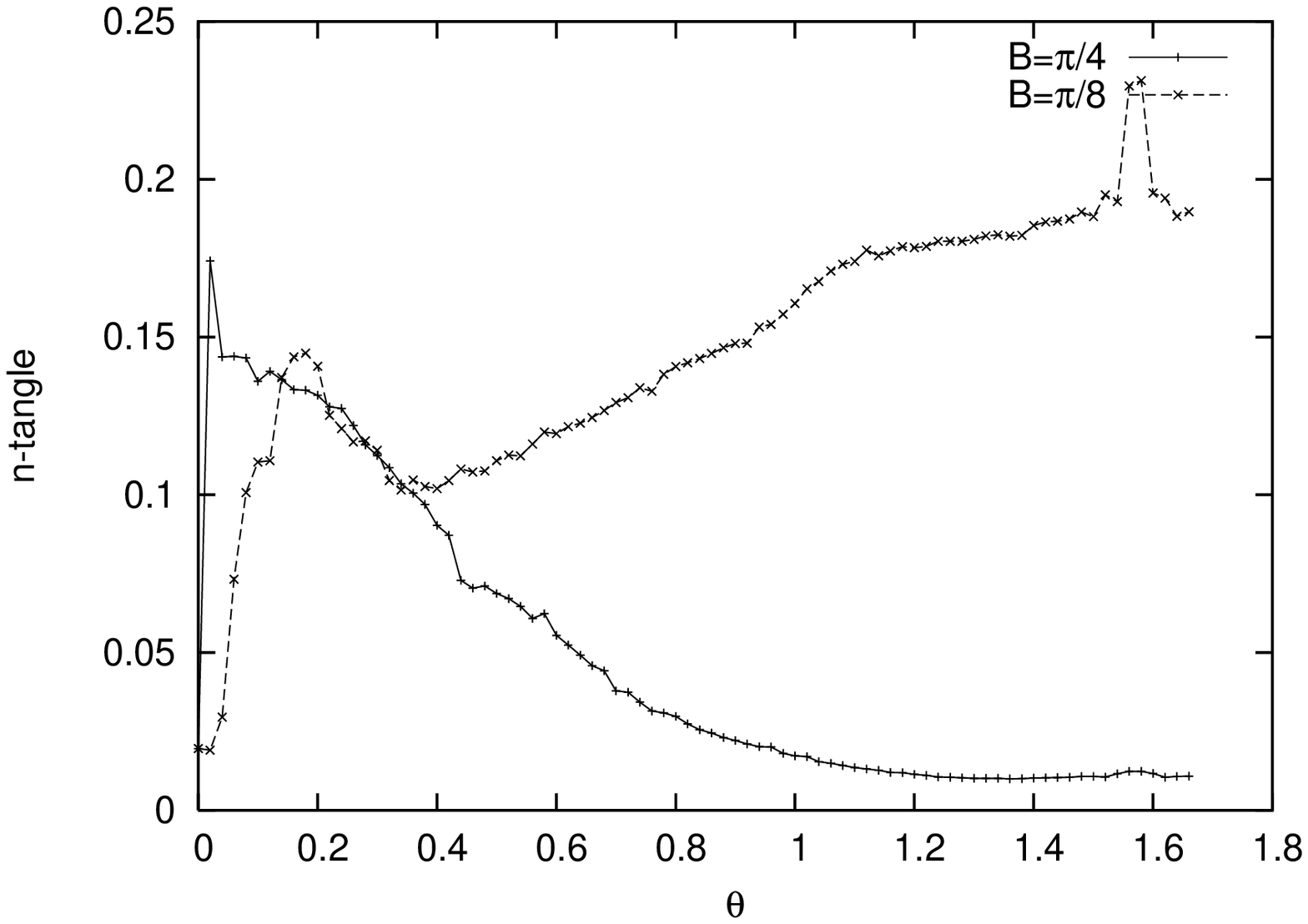}
\caption{The time averaged entanglement $Q$ and $n$-tangle as a function of the tilt angle, for two magnitudes of
the external field. The parameters are  $J_x=\pi/4$, $L=10$.}
\label{tilt5a6}
\end{figure*}

In order to see the effect of the angle and field strength more comprehensively, we again
time average the entanglement measures. This averaging is done over a large number ($1000$) of 
kicks such that the average is stationary. The results of this are shown in Fig.~(\ref{tilt5a6}),
where it is seen that the $Q$ measure increases sharply with the angle for a fixed magnitude
$B$ of the field, and then decreases smoothly till the transverse field is reached. The sharp
increase is observed in the case $B=J_x$, while smoother behavior is seen otherwise.
 The $n$-tangle measure shows similar characteristic, except that in one case the 
 transverse field case too has a high average entanglement value.

\begin{figure}
\includegraphics[height=3.5in,angle=0]{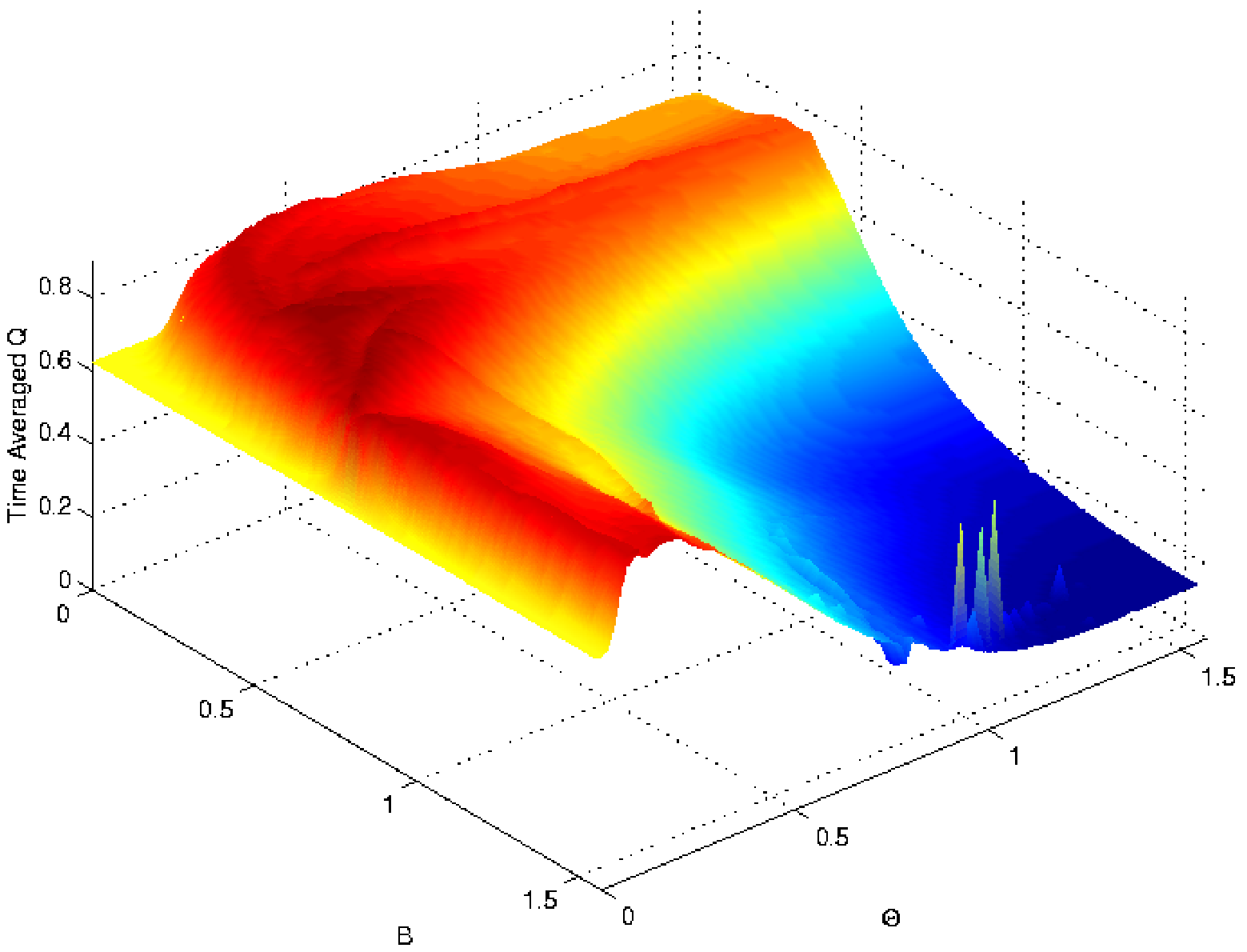}
\includegraphics[height=3.5in,angle=0]{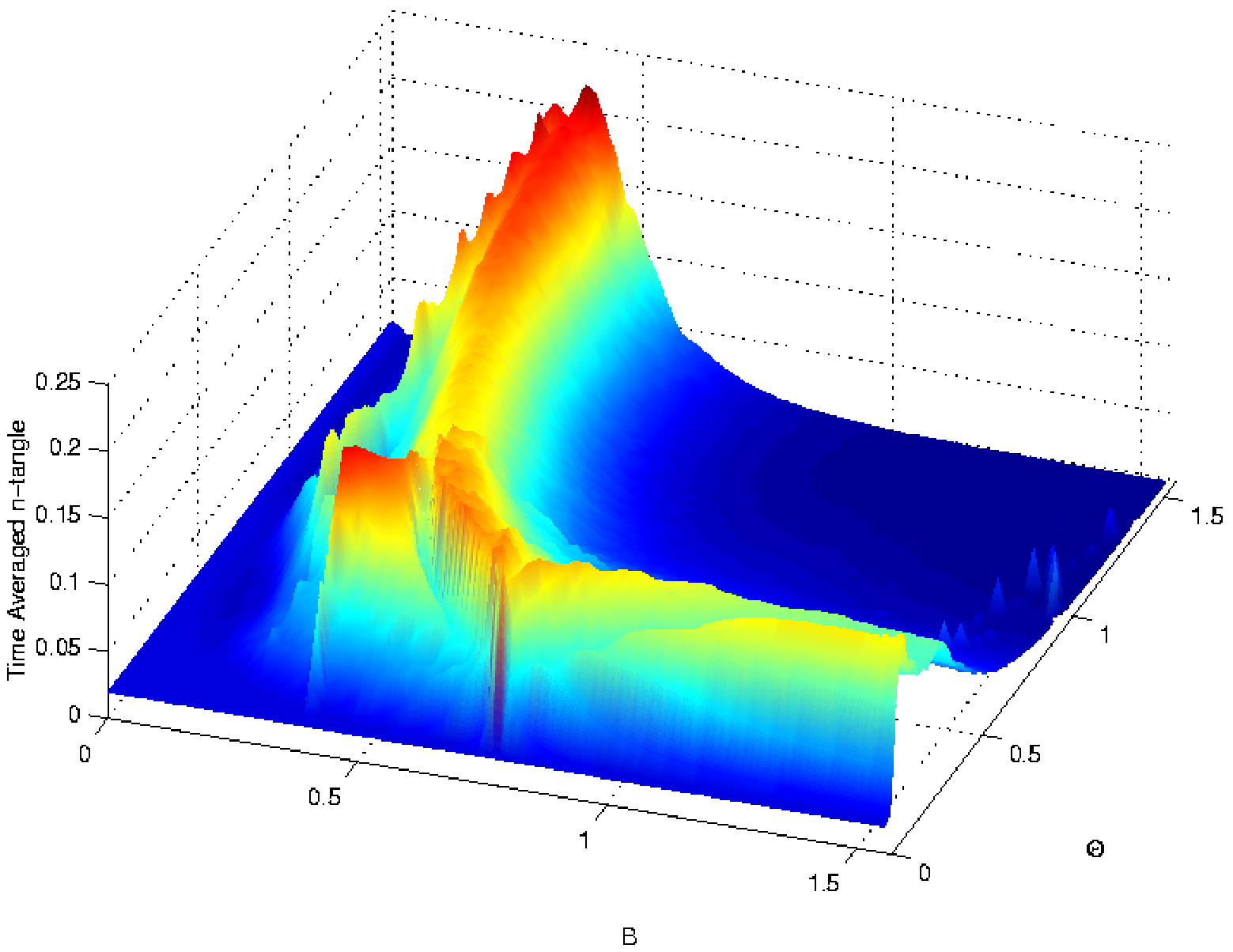}
\caption{(Color Online) The time averaged $Q$ and $n$-tangle as a function of external field parameters for the kicked transverse
Ising model. $J_x=\pi/4$ and $L=6$ in this case.}
\label{tiltavQnt}
\end{figure}

We next study the time averaged entanglement measures as a function of field strength and tilt, for a fixed
exchange coupling $J_x$. The averaging is done over large enough times to ensure stationarity
of this quantity, and is shown in Fig.~(\ref{tiltavQnt}). Only six spins are considered here 
as for each field configuration time evolution is done one thousand times, before calculating
the average. However the case of larger number of spins is qualitatively similar. The principal
features seen for the $Q$ measure is that there is enhanced entanglement for both small, nonzero,
field strengths and tilt angles. The sharp transition at $B=J_x$ is seen as a fold in the surface
plot of this figure. The high entanglement spots fall in approximate hyperbolas in the $B-\theta$ space.
The time-averaged $n$-tangle is also shown in Fig.~(\ref{tiltavQnt}), where the hyperbolic region of
high entanglement is also visible, but not so close to the small field and tilt angle values
as for the $Q$ measure.  
 
From the results presented so far it appears that entanglement can be enhanced in nonintegrable
regions of the spin chains, but there could be integrable regions such as for zero tilt angle
case which could produce large entanglement. We have not shown results for the residual tangle
in these cases, as this measure is {\em practically identical} to $Q$, this in turn implying that
the sum of the concurrences is nearly vanishing. In other words two-body correlations as measured
by the concurrence is a rare commodity in these spins chains. 
 More work needs to be done, especially with
different initial states, for a better understanding of the implications of nonintegrablity on
the entanglement in spin chains. The kicked transverse Ising model in a tilted field
is a natural example to explore this further.

\section{summary}

We have studied the $Q$, the $n$-tangle, the residual tangle and concurrence  measures for
a spin chain that is capable of showing both integrable and nonintegrable behaviours.
The model is the kicked Ising model, kicked with a field that could be transverse
or tilted to the exchange coupling direction. The integrable cases
correspond to the zero, parallel and transverse fields. In the zero or parallel cases
the states generated from the vacuum state are essentially the "cluster" states, 
for which we have derived the entanglement measures and shown that while the $Q$ measure is large,
the $n-$tangle measure can be exponentially small and the concurrences can vanish. We also
point out that symmetrization produces highly entangled states that are capable of both large
$Q$ and large $n$-tangles. 

In the case of the transverse field, we solve the time evolution by means of the Jordan-Wigner
transformation exactly. This enables calculation of many quantities analytically, of which 
we have displayed the $Q$ measure and pointed out the combinations of field strength
and exchange couplings that lead to states with large entanglement.
The Jordan-Wigner transformation does not help in the case of the 
tilted field and is an  nonintegrable case that has been previously studied
from a fidelity point of view. We have studied this case numerically 
and shown that time averaged entanglement can be enhanced in the nonintegrable cases, however it
is quite likely that this entanglement is not in the form of two-body entanglements.
A more detailed study of the nonintegrable case needs to be carried out to fully assess the
impact of nonitegrability on multipartite entanglement.

The entanglement measures $Q$ and the $n$-tangle have been calculated for random states 
and it has been shown using quantized chaotic maps that these are realized for states
evolving under conditions of quantum chaos \cite{ScottCaves,Scott}. The random state 
entanglement measure $Q$ for instance is an overestimate for the kicked Ising model
even with a tilted magnetic field, most likely indicating the effects of translation symmetry,
placing strong constraints on the ``randomness'' of these states. Future directions are many,
including a more detailed study of states that been shown here to have both large $Q$ and 
$n$-tangles, especially from an information theoretic viewpoint. Another, is the evaluation of
the issues studied here with other multipartite entanglement measures, for instance 
the distance to the nearest completely unentangled state \cite{Other}.

\begin{acknowledgments}
VS would like to thank The Institute of Mathematical Sciences, Chennai, for hospitality.
\end{acknowledgments}

\end{document}